\documentstyle[12pt]{article}
\begin{document}
\setlength{\baselineskip}{20pt}

\mbox{}

\begin{center}
{\Large{\bf New four-dimensional solutions of the Jacobi equations for Poisson structures}} \\ 
\mbox{} \\
{\bf Benito Hern\'{a}ndez--Bermejo$^{\; \mbox{\footnotesize {\rm a)}}}$}
\end{center}
{\em Departamento de Matem\'{a}ticas y F\'{\i}sica Aplicadas y Ciencias de la Naturaleza.} \\
{\em Universidad Rey Juan Carlos. Campus de M\'{o}stoles, Edificio Departamental II.} \\
{\em Calle Tulip\'{a}n S/N. 28933--M\'{o}stoles--Madrid. Spain.}

\mbox{}

\begin{center} 
{\bf Abstract}
\end{center}

A new four-dimensional family of skew-symmetric solutions of the Jacobi equations for 
Poisson structures is characterized. As a consequence, previously known types of Poisson structures found in a diversity of physical situations appear to be obtainable as particular cases of the new family of solutions. Additionally, it is possible to apply constructive methods for the explicit determination of fundamental properties of those solutions, such as their Casimir invariants, symplectic structure and the algorithm for the reduction to the Darboux canonical form, which have been reported only for a limited sample of known 
finite-dimensional Poisson structures. Moreover, the results developed are valid globally in phase space, thus ameliorating the usual scope of Darboux theorem which is of local nature. 

\mbox{}

\mbox{}



\noindent {\bf Running Title:} New solutions of the 4D Jacobi equations.

\mbox{}

\vfill

\mbox{}

\footnoterule
\noindent $^{\mbox{\footnotesize {\rm a)}}}$ E-mail: benito.hernandez@urjc.es 

\pagebreak
\begin{flushleft}
{\bf I. INTRODUCTION}
\end{flushleft}

Poisson structures$^{1,2\:}$ are present in many different domains of mathematical physics, 
such as fluid dynamics,$^{3\:}$ plasma physics,$^{4\:}$ field theory,$^{5\:}$ continuous 
media,$^{6\:}$ etc. In particular, finite-dimensional Poisson structures (to which this work is devoted) are relevant in the study of very different kinds of nonlinear systems, including population dynamics,$^{7-12\:}$ mechanics,$^{13-16\:}$ electromagnetism,$^{17\:}$ 
optics,$^{18\:}$ or plasma physics,$^{19\:}$ to cite a sample. The association of a 
finite-dimensional Poisson structure to a differential system (which is still an open 
problem$^{16,20,21,22\:}$) is not only mathematically appealing, but also very useful through the use of a plethora of specialized techniques which include the development of perturbative solutions,$^{17\:}$ numerical algorithms,$^{23\:}$ stability analysis by means of the 
energy-Casimir$^{24\:}$ and energy-momentum$^{25\:}$ methods, characterization of 
invariants,$^{26\:}$ reductions,$^{2,27\:}$ analysis of integrability properties,$^{28\:}$ establishment of variational principles,$^{29\:}$ study of bifurcation properties and chaotic behavior,$^{18,30\:}$ etc. 

When expressed in terms of a system of local coordinates on an $n$-dimensional manifold,  finite-dimensional Poisson structures take the form:
\begin{equation}
    \label{nham}
    \dot{x}_i = \sum_{j=1}^n J_{ij} \partial _j H \; , \;\:\; i = 1, \ldots , n
\end{equation} 
Here and in what follows $ \partial_j \equiv \partial / \partial x_j$. The $C^1$ real-valued function $H(x)$ in (\ref{nham}) is a constant of motion of the system playing the role of Hamiltonian. The $J_{ij}(x)$, called structure functions, are also $C^1$ and real-valued and constitute the entries of an $n \times n$ structure matrix ${\cal J}$. The $J_{ij}(x)$ are 
characterized by two properties. The first one is that they are skew-symmetric:
\begin{equation}
      \label{sksym}
	J_{ij}=-J_{ji} \;\:\;\: \mbox{for all} \; \: i,j 
\end{equation}
And second, they are solutions of the Jacobi equations
\begin{equation}
     \label{jac}
     \sum_{l=1}^n ( J_{il} \partial_l J_{jk} + J_{kl} \partial_l J_{ij} + 
	J_{jl} \partial_l J_{ki} ) = 0 
\end{equation}
where indices $i,j,k$ run from 1 to $n$ in equations (\ref{sksym}) and (\ref{jac}). 

One of the reasons justifying the importance of the Poisson representation is the local 
equivalence bewteen Poisson systems and classical Hamiltonian systems, as stated by Darboux 
theorem$^{1,2}$ which demonstrates that if an $n$-dimensional Poisson manifold has constant rank of value $2r$ everywhere, then at each point of the manifold there exist local coordinates $(p_1, \ldots ,p_{r},q_1, \ldots , q_{r},z_1, \ldots , z_{n-2r})$ in terms of which the equations of motion become:
\[
	\dot{q}_i = \frac{\partial H}{\partial p_i} \;\: , \;\:\:
	\dot{p}_i = - \frac{\partial H}{\partial q_i} \;\: , \;\:\: i=1, \ldots ,r 
\]

\[
	\dot{z}_j = 0 \;\: , \;\:\: j=1 , \ldots , n-2r
\]

As mentioned above, the problem of recasting a given vector field not explicitly written in the form (\ref{nham}) in terms of a finite-dimensional Poisson system is an open issue of fundamental importance in this context to which important efforts have been devoted in past years in a variety of approaches and situations.$^{7-22\:}$ This explains, together with the intrinsic mathematical interest of the problem, the permanent attention deserved in the literature by the obtainment and classification of skew-symmetric solutions of the Jacobi equations.$^{7-22,31-38\:}$ Given that equations (\ref{jac}) constitute a set of coupled nonlinear partial differential equations, the characterization of solutions of 
(\ref{sksym}-\ref{jac}) has proceeded by means of either suitable {\em ansatzs\/}$^{7-11,32,37\:}$ or through a diversity of other approaches.$^{12-16,20-22,31,38\:}$ These efforts have led to the determination of certain families of solutions of increasing nonlinearity such as the constant ones (of which the symplectic matrices are just a particular case), as well as linear$^{2,33\:}$ (i.e. Lie-Poisson), affine-linear,$^{34\:}$ quadratic,$^{7-11,15,35,36\:}$ and cubic$^{37\:}$ structures, together with solutions which comprise arbitrary 
functions.$^{12-14,16,20-22,31,32,38\:}$ Simultaneously, the growing complexity of the Jacobi equations (\ref{jac}) as the dimension $n$ increases has determined that the analysis is often focused on three-dimensional solutions,$^{9,10,12,20,21,32,37,38\:}$ while the characterization of families of dimensions four,$^{13\:}$ five,$^{14\:}$ six,$^{17\:}$ etc. is less frequent. In addition, some wide families of $n$-dimensional solutions have also been analyzed in the literature.$^{8,11,31,33-36\:}$ 

In this work a new four-dimensional family of solutions of the Jacobi equations (\ref{jac}) is characterized. This contribution presents several interesting features. First, it is worth noting that previously known types of Poisson structures appearing in a diversity of physical situations and systems can be seen to be obtainable as particular cases of the new family of solutions, as it will be seen in the examples section. Second, in spite of their generality the solutions to be considered in what follows are amenable to explicit and detailed analysis, since it is possible to characterize globally their Casimir invariants and symplectic structure, as well as to globally provide the reduction to the Darboux canonical form. This constitutes a significant amelioration of the usual scope of Darboux theorem, which does only guarantee in principle a local reduction.$^{1,2\:}$ In addition, the achievement of such reduction is relevant as far as the explicit determination of the Darboux coordinates is often a complicated task, only known for a limited sample of finite-dimensional Poisson structures.$^{2,8,27,31,38\:}$ 

The structure of the article is as follows. In Section II the new solutions are characterized. 
The symplectic structure and the constructive reduction to the Darboux canonical form are investigated in Section III. Examples and comments on the relationship with some previously known results are provided in Section IV. The work concludes in Section V with some final remarks. 

\mbox{}

\begin{flushleft}
{\bf II. CHARACTERIZATION OF THE FAMILY OF SOLUTIONS}
\end{flushleft}

We begin with one of the main results: 

\mbox{}

{\bf Theorem 2.1: \/} Consider the family of functions of the form
\begin{equation}
\label{sol}
J_{ij}(x)= \sigma _{ij} \eta (x) \psi _i (x_i) \psi _j (x_j) \sum_{k,l=1}^4 \epsilon _{ijkl} 
\phi _l (x_l) \:\: , \:\:\: i,j = 1, \ldots , 4
\end{equation}
defined in an open domain $\Omega \subset I \!\! R^4$, where $\epsilon _{ijkl}$ denotes the Levi-Civita symbol and such that:
\begin{description}
\item[{\rm (a)}] Constants $\sigma _{ij} \in I \!\! R$ are defined for every pair $(i,j)$, $i \neq j$. 
\item[{\rm (b)}] $\sigma _{ij}= \sigma _{ji}$ for every pair $(i,j)$, $i \neq j$.
\item[{\rm (c)}] $\sigma _{ij} \neq 0$ for at least one pair $(i,j)$, $i \neq j$.
\item[{\rm (d)}] $\eta(x)$, $\psi_i(x_i)$ and $\phi_i(x_i)$ are $C^1(\Omega)$ functions of their respective arguments for every $i$. 
\item[{\rm (e)}] $\eta (x)$ and $\psi _i(x_i)$ are nonvanishing in $\Omega$ for every $i$.
\item[{\rm (f)}] The differences $( \phi _i(x_i)- \phi _j(x_j))$ are nonvanishing in $\Omega$ 
for every pair $(i,j)$, $i \neq j$.
\end{description}
Then the set of functions $J_{ij}(x)$ defined in (\ref{sol}) constitutes a skew-symmetric solution of the four-dimensional Jacobi identities 
\begin{equation}
     \label{jac4d}
     \sum_{l=1}^4 ( J_{il} \partial_l J_{jk} + J_{kl} \partial_l J_{ij} + 
	J_{jl} \partial_l J_{ki} ) = 0 \:\: , \:\:\:\:\: i,j,k=1, \ldots ,4
\end{equation}
and therefore ${\cal J} = (J_{ij})$ is a four-dimensional structure matrix, if and only if: 
\begin{equation}
\label{condsig}
	\sigma_{12}\sigma_{34}=\sigma_{13}\sigma_{24}=\sigma_{14}\sigma_{23}
\end{equation}

\mbox{}

{\em Proof: \/} Consider first functions (\ref{sol}) in the case $\eta =1$. Substitution of (\ref{sol}) in equation (\ref{jac4d}) of indexes $(i,j,k)$ leads after some algebra to:
\begin{equation}
\label{lbg}
	\begin{array}{c}
	\sum_{l=1}^4 ( J_{il} \partial_l J_{jk} + J_{kl} \partial_l J_{ij} + 
	J_{jl} \partial_l J_{ki} ) = \\ \\
	\psi _i \psi _j \psi _k \sum_{r_1,r_2,s_1,s_2=1}^4 \left\{ 
	(\sigma _{ij} \sigma _{jk} \epsilon _{ijr_1r_2} \epsilon _{jks_1s_2} + 
	\sigma _{kj} \sigma _{ij} \epsilon _{kjr_1r_2} \epsilon _{ijs_1s_2} )
	(\partial_j \psi_j) \phi_{r_2} \phi_{s_2} + \right. \\ \\
	(\sigma _{ki} \sigma _{ij} \epsilon _{kir_1r_2} \epsilon _{ijs_1s_2} + 
	\sigma _{ji} \sigma _{ki} \epsilon _{jir_1r_2} \epsilon _{kis_1s_2} )
	(\partial_i \psi_i) \phi_{r_2} \phi_{s_2} + \\ \\
	(\sigma _{ik} \sigma _{jk} \epsilon _{ikr_1r_2} \epsilon _{jks_1s_2} + 
	\sigma _{jk} \sigma _{ki} \epsilon _{jkr_1r_2} \epsilon _{kis_1s_2} )
	(\partial_k \psi_k) \phi_{r_2} \phi_{s_2} + \\ \\ \left.
	(\sigma _{is_2} \sigma _{jk} \epsilon _{is_2r_1r_2} \epsilon _{jks_1s_2}+
	\sigma _{ks_2} \sigma _{ij} \epsilon _{ks_2r_1r_2} \epsilon _{ijs_1s_2}+
	\sigma _{js_2} \sigma _{ki} \epsilon _{js_2r_1r_2} \epsilon _{kis_1s_2}) 
	\psi_{s_2} \phi_{r_2} (\partial_{s_2} \phi_{s_2}) \right\} = \\ \\
	\psi _i \psi _j \psi _k \sum_{r_1,r_2,s_1,s_2=1}^4 \left\{ (
	\sigma_{is_2}\sigma_{jk} \delta^{is_2r_1r_2}_{jks_1s_2} +
	\sigma_{ks_2}\sigma_{ij} \delta^{ks_2r_1r_2}_{ijs_1s_2} +
	\sigma_{js_2}\sigma_{ki} \delta^{js_2r_1r_2}_{kis_1s_2} ) 
	\psi_{s_2} \phi_{r_2} (\partial_{s_2} \phi_{s_2}) \right\} 
	\end{array}
\end{equation}
where the $\delta$ symbol denotes the generalized Kronecker delta according to its standard definition, namely: given $q$ superindexes $(i_1, \ldots , i_q)$ and $q$ subindexes $(j_1,\ldots , j_q)$ all of them taking values in the range $(1, \ldots ,n)$, then $\delta ^{i_1 \ldots i_q}_{j_1 \ldots j_q}$ is defined by the properties: (a) it is totally antisymmetric in the superindexes; (b) it is totally antisymmetric in the subindexes; (c) if the superindexes are all different (this is, $i_{a_1} \neq i_{a_2}$ if $a_1 \neq a_2$) and the subindexes are a permutation of the superindexes, then $\delta ^{i_1 \ldots i_q}_{j_1 \ldots j_q}$ takes the value $+1$ (respectively, $-1$) if $(i_1, \ldots , i_q)$ and $(j_1,\ldots , j_q)$ are permutations of the same (of different) sign; (d) the value of $\delta ^{i_1 \ldots i_q}_{j_1 \ldots j_q}$ is zero otherwise. Consequently, it can be verified that the expression in (\ref{lbg}) vanishes if two of the three indexes $(i,j,k)$ are equal. Consider then the case in which $i$, $j$ and $k$ are different. If $m$ is the integer, $1 \leq m \leq 4$, such that $(i,j,k,m)$ is a permutation of $(1,2,3,4)$, we arrive at:
\begin{equation}
\label{ijkdifs}
	\begin{array}{c}
	\sum_{l=1}^4 ( J_{il} \partial_l J_{jk} + J_{kl} \partial_l J_{ij} + 
	J_{jl} \partial_l J_{ki} ) = \\ \\
	\psi_i \psi_j \psi_k \psi_m (\partial_m \phi_m) \left\{ 
	\sigma_{im} \sigma_{jk}(\phi_k-\phi_j) + \sigma_{km} \sigma_{ij}(\phi_j-\phi_i) +
	\sigma_{jm} \sigma_{ki}(\phi_i-\phi_k) \right\} = \\ \\
	\psi_i \psi_j \psi_k \psi_m (\partial_m \phi_m) \left\{ 
	(\sigma_{jm} \sigma_{ki} - \sigma_{km} \sigma_{ij}) \phi_i +
	(\sigma_{km} \sigma_{ij} - \sigma_{im} \sigma_{jk}) \phi_j +
	(\sigma_{im} \sigma_{jk} - \sigma_{jm} \sigma_{ki}) \phi_k \right\} 
	\end{array}
\end{equation}
Now let $p$, where $0 \leq p \leq 4$, be the number of functions $\phi_i$ which have constant value everywhere in $\Omega$. Taking into account hypothesis (f) of the theorem, there are five different possibilities to be examined for equation (\ref{ijkdifs}):
\begin{description}
\item[$p=0:$] in this case it is straightforward that (\ref{ijkdifs}) vanishes if and only if 
(\ref{condsig}) holds. 
\item[$p=1:$] the analysis and the result are similar to those of the case $p=0$.
\item[$p=2:$] assume without loss of generality that $\phi_k$ and $\phi_m$ 
are constant in $\Omega$ while $\phi_i$ and $\phi_j$ are not. Then expression (\ref{ijkdifs}) vanishes if and only if:
\[
	\sigma_{im} \sigma_{jk} - \sigma_{jm} \sigma_{ik} = 
	(\sigma_{im} \sigma_{jk}- \sigma_{ij} \sigma_{km}) \phi_k +(\sigma_{ij} \sigma_{km} - 	\sigma_{im} \sigma_{jk}) \phi_m =0
\]
Given that $\phi_k \neq \phi_m$, these equations are equivalent to (\ref{condsig}).
\item[$p=3:$] suppose without loss of generality that $\phi_i$, $\phi_j$ and $\phi_k$ are constant in $\Omega$, while $\phi_m$ is not. Then expression (\ref{ijkdifs}) is equal to zero if and only if:
\[
	\left\{ (\sigma_{jm} \sigma_{ki}- \sigma_{km} \sigma_{ij}) \phi_i +
	(\sigma_{km} \sigma_{ij} - \sigma_{im} \sigma_{jk}) \phi_j +
	(\sigma_{im} \sigma_{jk}- \sigma_{jm} \sigma_{ki}) \phi_k \right\} \partial_m \phi_m = 0  
\]
Taking into account that $\partial_m \phi_m$ does not vanish everywhere in $\Omega$, and that $\phi_i$, $\phi_j$ and $\phi_k$ are arbitrary (as far as hypothesis (f) of the theorem is respected) the outcome is again that (\ref{condsig}) is necessary and sufficient for the vanishing of (\ref{ijkdifs}).
\item[$p=4:$] equations (\ref{ijkdifs}) vanish because $\partial_m \phi_m = 0$ for all possible values of $m$. This is to be expected because in this case we are dealing with a separable structure.$^{31}$
\end{description}
Then conditions (\ref{condsig}) are necesary and sufficient for the vanishing of 
(\ref{ijkdifs}) when $0 \leq p \leq 3$. For $p=4$ expression (\ref{ijkdifs}) is always zero. This concludes the analysis of the case $\eta =1$. 

\noindent Let us now turn to the general form (\ref{sol}) of the solution, namely to general $\eta $. To analyze this case, consider an arbitrary four-dimensional skew-symmetric solution $J_{ij}(x)$ of the Jacobi equations. If such solution is multiplied by a $C^1(\Omega)$ function $\eta (x)$ the resulting set of functions $J^*_{ij}(x) = \eta (x) J_{ij}(x)$ will be a skew-symmetric solution of (\ref{jac4d}) if and only if $\eta$ verifies:
\begin{equation}
\label{defor}
	\left( J_{im}J_{jk} + J_{km}J_{ij} + J_{jm}J_{ki} \right) \partial _m \eta =0 
\end{equation}
where again $(i,j,k,m)$ denotes every permutation of $(1,2,3,4)$. We now apply condition 
(\ref{defor}) to the functions $J_{ij}$ in (\ref{sol}) for which $\eta =1$, just considered in the first part of this proof. It can thus be seen that:
\begin{equation}
\label{ecp10}
	\begin{array}{c}
	J_{im}J_{jk} + J_{km}J_{ij} + J_{jm}J_{ki} = \\ \\
	\psi_i \psi_j \psi_k \psi_m \sum_{p,q,r,s=1}^4 \phi_q \phi_s \left\{ 
	\sigma_{im} \sigma_{jk} \delta _{impq}^{jkrs} + 
	\sigma_{ij} \sigma_{km} \delta _{kmpq}^{ijrs} + 
	\sigma_{jm} \sigma_{ki} \delta _{jmpq}^{kirs} \right\}
	\end{array}
\end{equation}
\noindent To evaluate this expression, consider first the cases $0 \leq p \leq 3$, which are verified if and only if (\ref{condsig}) is valid. In such situations equation (\ref{ecp10}) becomes
\[
	\begin{array}{c}
	J_{im}J_{jk} + J_{km}J_{ij} + J_{jm}J_{ki} = \\ \\
	\psi_i \psi_j \psi_k \psi_m \sigma_{im} \sigma_{jk} \sum_{p,q,r,s=1}^4 \phi_q \phi_s 
	\left\{ \delta _{impq}^{jkrs} + \delta _{kmpq}^{ijrs} + \delta _{jmpq}^{kirs} \right\} =0
	\end{array}
\]
and the result is demonstrated. For the remaining case $p=4$ it can be seen after some algebra that (\ref{ecp10}) amounts to:
\begin{equation}
\label{casep0}
	\begin{array}{c}
	J_{im}J_{jk} + J_{km}J_{ij} + J_{jm}J_{ki} = 
	\psi_i \psi_j \psi_k \psi_m \left\{ 
	( \sigma_{im} \sigma_{jk} - \sigma_{jm} \sigma_{ki}) (\phi_i \phi_j + \phi_k \phi_m)+ 
	\right. \\ \\ \left.
	( \sigma_{ij} \sigma_{km} - \sigma_{im} \sigma_{jk}) (\phi_i \phi_k + \phi_j \phi_m)+
	( \sigma_{jm} \sigma_{ki} - \sigma_{ij} \sigma_{km}) (\phi_i \phi_m + \phi_j \phi_k)
	\right\}
	\end{array}
\end{equation}
This expression must vanish everywhere in $\Omega$ if (\ref{sol}) is to be a solution for arbitrary $\eta$ in this case. Since $p=4$ (namely all $\phi_i$ are constant in $\Omega$) then hypothesis (f) implies that there are two possibilities: either $\phi_i \neq 0$ for every $i=1, \ldots ,4$; or $\phi_i = 0$ for just one value of $i$. It can be shown in both situations that (\ref{casep0}) vanishes if and only if (\ref{condsig}) is verified. Consequently, the inclusion of function $\eta$ implies that conditions (\ref{condsig}) are also necessary and sufficient in the case $p=4$. This completes the proof of Theorem 2.1. \hfill Q.E.D.

\mbox{}

Therefore the family of Poisson structures just characterized has the matrix form
\begin{equation}
\label{matrform}
	{\cal J} = \eta \cdot
	\left( \begin{array}{cccc}
	 0 & \sigma_{12} \psi_1 \psi_2 (\phi_4 - \phi_3) 
	& \sigma_{13} \psi_1 \psi_3 (\phi_2 - \phi_4) 
	& \sigma_{14} \psi_1 \psi_4 (\phi_3 - \phi_2) \\
	\sigma_{12} \psi_1 \psi_2 (\phi_3 - \phi_4) & 0 &
	\sigma_{23} \psi_2 \psi_3 (\phi_4 - \phi_1) &
	\sigma_{24} \psi_2 \psi_4 (\phi_1 - \phi_3) \\
	\sigma_{13} \psi_1 \psi_3 (\phi_4 - \phi_2) & 
	\sigma_{23} \psi_2 \psi_3 (\phi_1 - \phi_4) & 0 &
	\sigma_{34} \psi_3 \psi_4 (\phi_2 - \phi_1) \\
	\sigma_{14} \psi_1 \psi_4 (\phi_2 - \phi_3) & 
	\sigma_{24} \psi_2 \psi_4 (\phi_3 - \phi_1) &
	\sigma_{34} \psi_3 \psi_4 (\phi_1 - \phi_2) & 0 \\	
\end{array} \right)
\end{equation}
where additionally $\sigma_{12} \sigma_{34} = \sigma_{13} \sigma_{24} = \sigma_{14} \sigma_{23}$. For what is to follow, the next definition will be necessary:

\mbox{}

{\em Definition 2.2: \/} For every open domain $\Omega \subset I \! \! R^4$, the set of 
Poisson structures defined in $\Omega$ and of the kind (\ref{sol}) characterized in Theorem 2.1 will be denoted $\Theta ( \Omega )$.

\mbox{}

To provide the basis for the analysis of the symplectic structure and Darboux reduction in Section III and also in order to complete the description of these structures, the following result is important:

\mbox{}

{\em Proposition 2.3: \/} Let $\Omega \subset I \! \! R^4$ be an open set, then every Poisson structure ${\cal J} \in \Theta ( \Omega )$ has constant rank of value 2 everywhere in $\Omega$.

\mbox{}

{\em Proof: \/} The determinant of ${\cal J}$ in (\ref{matrform}) is:
\[
	\begin{array}{c}
	\mid {\cal J} \mid = \eta (\psi_1 \psi_2 \psi_3 \psi_4 )^2 [
	(\sigma_{14} \sigma_{23} - \sigma_{13} \sigma_{24}) ( \phi_1 \phi_2 + \phi_3 \phi_4)+ 
	\\ \\
	(\sigma_{12} \sigma_{34} - \sigma_{14} \sigma_{23}) ( \phi_1 \phi_3 + \phi_2 \phi_4)+
	(\sigma_{13} \sigma_{24} - \sigma_{12} \sigma_{34}) ( \phi_1 \phi_4 + \phi_2 \phi_3)]^2
	\end{array}
\]
Due to identities (\ref{condsig}) the result is that $\mid {\cal J} \mid = 0$. Therefore the rank cannot be 4, but only 2 or 0. The fact that the rank is 2 everywhere in $\Omega$ is implied by conditions (c), (e) and (f) of Theorem 2.1. \hfill Q.E.D.

\mbox{}

Proposition 2.3 provides the basis for the explicit determination of the symplectic structure and Darboux reduction of these structures. This is the purpose of the next section. 

\mbox{}

\begin{flushleft}
{\bf III. SYMPLECTIC STRUCTURE AND DARBOUX CANONICAL FORM}
\end{flushleft}

Before developing the main issues of this section it is necessary to recall a known definition$^{38\:}$ that will be needed for their establishment:

\mbox{}

{\em Definition 3.1: \/} Let $\Omega \subset I \!\! R^4$ be an open set. A reparametrization 
of time is defined as a transformation of the form
\begin{equation}
	\label{ntt}
	\mbox{d}\tau = \frac{1}{\mu(x)}\mbox{d}t
\end{equation}
where $t$ is the initial time variable, $\tau$ is the new time and $\mu (x) : \Omega 
\longrightarrow I \!\! R$ is a $C^1(\Omega)$ function which does not vanish in $\Omega$.

\mbox{}

The sense of this definition is the following: let
\begin{equation}
	\label{4dpos}
	\frac{\mbox{d}x}{\mbox{d}t} = {\cal J} \cdot \nabla H
\end{equation}
be an arbitrary four-dimensional Poisson structure defined in an open domain 
$\Omega \subset I \!\! R^4$. Then, every reparametrization of time of the form (\ref{ntt}) 
leads from (\ref{4dpos}) to the differential system:
\begin{equation}
	\label{4dposntt}
	\frac{\mbox{d}x}{\mbox{d} \tau} = \mu {\cal J} \cdot \nabla H
\end{equation}
Note however that such transformation often destroys the Poisson structure for systems of dimension higher than three,$^{38\:}$ because for a given ${\cal J}$ which is a structure matrix, $\mu {\cal J}$ is not necessarily a solution of (\ref{sksym}-\ref{jac}) as it has been discussed in the proof of Theorem 2.1 in connection with the four-dimensional case.

The main purpose of this section is the investigation of the symplectic structure of family $\Theta ( \Omega )$. The central result in this sense corresponds to the next theorem, for which the proof is constructive and completely classifies the different cases arising in the explicit determination of the Casimir invariants and the global reduction to the Darboux canonical form for the members of $\Theta ( \Omega )$:

\mbox{}

{\bf Theorem 3.2: \/} For every four-dimensional Poisson system
\[
	\frac{\mbox{d}x}{\mbox{d} t} = {\cal J} \cdot \nabla H
\]
defined in an open domain $\Omega \subset I \!\! R^4$ and such that ${\cal J} \in \Theta 
( \Omega )$, both a complete set of $C^2( \Omega )$ independent Casimir invariants as well as  the reduction to the Darboux canonical form, can be globally constructed in $\Omega$.

\mbox{}

{\em Proof: \/} The proof begins with an auxiliary result:

\mbox{}

{\em Lemma 3.3: \/} Let $\Omega \subset I \!\! R^4$ be an open set, then every ${\cal J} \in \Theta ( \Omega )$ is equivalent to a Poisson structure ${\cal J}'$ defined in a domain $\Omega '$, of rank constant and equal to 2 in $\Omega '$ and components of the form
\begin{equation}
\label{red1}
	J'_{ij}(y)= \sigma _{ij} \eta '(y)  \sum_{k,l=1}^4 \epsilon _{ijkl} \phi '_l (y_l) 
	\:\: , \:\:\: i,j = 1, \ldots , 4
\end{equation}
Moreover, ${\cal J}'$ is obtained through the change of variables globally diffeomorphic in $\Omega$
\begin{equation}
	\label{yxdiff}
      y_i(x_i) = \int \frac{\mbox{d}x_i}{\psi_i (x_i)} \:\: , \:\:\:\:\: i=1, \ldots ,4
\end{equation}
and $\Omega ' = y( \Omega )$ is the diffeomorphic image of $\Omega$ through transformation (\ref{yxdiff}).

\mbox{}

{\em Proof of Lemma 3.3: \/} Recall that after a general diffeomorphism $y = y(x)$, a given structure matrix ${\cal J}(x)$ is transformed into another one ${\cal J'}(y)$ according to the  rule:
\begin{equation}
	\label{jdiff}
      J'_{ij}(y) = \sum_{k,l=1}^n \frac{\partial y_i}{\partial x_k} J_{kl}(x) 
	\frac{\partial y_j}{\partial x_l}
\end{equation}
The use of (\ref{jdiff}) with transformation (\ref{yxdiff}) on ${\cal J}$ leads to (\ref{red1}) with $\eta '(y) = \eta (x(y))$ and $\phi'_i(y) = \phi_i(x(y))$ for $i=1, \ldots ,4$. The fact that the rank of (\ref{red1}) is constant and of value 2 everywhere in $\Omega '$ is a direct consequence of Proposition 2.3 and identity (\ref{jdiff}). \hfill Q.E.D.

\mbox{}

The Poisson structure (\ref{red1}) will be the starting point for the rest of the proof. Now two complementary cases are to be distinguished:

\begin{description}
\item[{\rm CASE I:}] $\sigma_{ij} \neq 0$ for all pairs $(i,j)$, $i \neq j$. The analysis of this case must begin with a definition and some preliminary results:

\mbox{}

{\em Definition 3.4: \/} Given an open set $\Omega \subset I \! \! R^4$, a Poisson structure belonging to $\Theta (\Omega)$ is said to be $\sigma$-positive if all its constants $\sigma_{ij}$ can be chosen to be positive, where $i,j=1, \ldots ,4$ and $i \neq j$.

\mbox{}

{\em Lemma 3.5: \/} Let $\Omega \subset I \! \! R^4$ be an open set, and let ${\cal{J}} \in \Theta (\Omega)$ be a Poisson structure for which $\sigma_{ij} \neq 0$ for every pair $i \neq j$, where $i,j=1, \ldots ,4$. Then $\cal{J}$ is $\sigma$-positive and can be expressed in terms of the set of constants $\tilde{\sigma}_{ij}= \mid \sigma_{ij} \mid$.

\mbox{}

{\em Proof of Lemma 3.5: \/} From now on, we define $\sigma \equiv \sigma_{12}\sigma_{34}= \sigma_{13}\sigma_{24}= \sigma_{14}\sigma_{23}$ (recall equation (\ref{condsig})). Four main cases can be distinguished:
\begin{description}
\item[{\rm Case 1:}] $\sigma_{ij}>0$ for all $i \neq j$. The matrix is already in $\sigma$-positive form.
\item[{\rm Case 2:}] $\sigma_{ij}<0$ for all $i \neq j$. This is reduced to Case 1 by redefining $\phi_i(x_i)$ as $\tilde{\phi}_i(x_i) = -\phi_i(x_i)$ for every $i$.
\item[{\rm Case 3:}] $\sigma >0$ with constants $\sigma_{ij}$ both positive and negative. There are two subcases:
\begin{description}
\item[{\rm Case 3.1:}] There are two negative and four positive constants $\sigma_{ij}$ with $i<j$. 
\begin{description}
\item[{\rm Case 3.1.1:}] $\sigma_{12}<0$ and $\sigma_{34}<0$. 
\item[{\rm Case 3.1.2:}] $\sigma_{13}<0$ and $\sigma_{24}<0$.  
\item[{\rm Case 3.1.3:}] $\sigma_{14}<0$ and $\sigma_{23}<0$.  

	The three subcases 3.1.x are reduced in two steps: 
	\begin{description}
	\item[{\rm Step 1:}] redefine $\phi_i(x_i)$ as $\tilde{\phi}_i(x_i) = -\phi_i(x_i)$ for 
	every $i$.
	\item[{\rm Step 2:}] redefine $\psi_i(x_i)$ as $\tilde{\psi}_i(x_i) = -\psi_i(x_i)$ for 
	$i=3,4$ in subcase 3.1.1, for $i=1,3$ in subcase 3.1.2 and for $i=1,4$ in subcase 3.1.3.
	\end{description}
\end{description}
\item[{\rm Case 3.2:}] There are two positive and four negative constants $\sigma_{ij}$ with $i<j$. These are three possible cases that coincide with the ones appearing after Step 1 of items 3.1.1, 3.1.2 and 3.1.3 and therefore their reduction corresponds to the transformations indicated in Step 2 of those three subcases.
\end{description}
\item[{\rm Case 4:}] $\sigma <0$. Clearly it can be assumed without loss of generality that $\sigma_{12}<0$. Then there are four possibilities:
\begin{description}
\item[{\rm Case 4.1:}] $\sigma_{13}<0$ and $\sigma_{14}<0$. Redefining 
$\tilde{\psi}_1(x_1) = -\psi_1(x_1)$ it is reduced to Case 1.
\item[{\rm Case 4.2:}] $\sigma_{13}>0$ and $\sigma_{14}>0$. Redefining 
$\tilde{\psi}_2(x_2) = -\psi_2(x_2)$ it is reduced to Case 1.
\item[{\rm Case 4.3:}] $\sigma_{13}>0$ and $\sigma_{14}<0$. Redefining 
$\tilde{\psi}_3(x_3) = -\psi_3(x_3)$ it is reduced to Case 2.
\item[{\rm Case 4.4:}] $\sigma_{13}<0$ and $\sigma_{14}>0$. Redefining 
$\tilde{\psi}_4(x_4) = -\psi_4(x_4)$ it is reduced to Case 2.
\end{description}
\end{description}
This completes the proof of Lemma 3.5. \hfill Q.E.D.

\mbox{}

A result that complements the last lemma is the next one:

\mbox{}

{\em Lemma 3.6: \/} For every set of positive real constants $\{ \sigma_{12}, 
\sigma_{13}, \sigma_{14}, \sigma_{23}, \sigma_{24}, \sigma_{34} \}$ verifying conditions 
(\ref{condsig}) there exists a unique set of positive real constants $\{ \sigma_{1}, 
\sigma_{2}, \sigma_{3}, \sigma_{4} \}$ such that the equalities $\sigma _{ij} = 
\sigma_i \sigma_j$ are satisfied for every pair $(i,j)$, with $i<j$, $1 \leq i \leq 3$, $2 \leq j \leq 4$.

\mbox{}

{\em Proof of Lemma 3.6: \/} The existence of the constants $\sigma_i$ can be seen on their explicit expressions 
\[
\sigma_1 = \left( \frac{\sigma_{12}\sigma_{13}\sigma_{14}}{ \sigma} \right)^{1/2} , \;\;\; 
\sigma_2 = \left( \frac{\sigma \sigma_{12}}{\sigma_{13}\sigma_{14}} \right)^{1/2} , \;\;\; 
\sigma_3 = \left( \frac{\sigma \sigma_{13}}{\sigma_{12}\sigma_{14}} \right)^{1/2} , \;\;\; 
\sigma_4 = \left( \frac{\sigma \sigma_{14}}{\sigma_{12}\sigma_{13}} \right)^{1/2} 
\]
where now $\sigma >0$. To prove uniqueness, taking logarithms of equalities $\sigma _{ij} = \sigma_i \sigma_j$ allows reducing the problem to the investigation of the following linear system:
\begin{equation}
\label{syslin}
\left( \begin{array}{cccc} 1 & 1 & 0 & 0 \\ 1 & 0 & 1 & 0 \\ 1 & 0 & 0 & 1 \\ 
0 & 1 & 1 & 0 \\ 0 & 1 & 0 & 1 \\ 0 & 0 & 1 & 1 \end{array} \right) \cdot 
\left( \begin{array}{c} \ln \sigma_1 \\ \ln \sigma_2 \\ \ln \sigma_3 \\
\ln \sigma_4 \end{array} \right) =
\left( \begin{array}{c} \ln \sigma_{12} \\ \ln \sigma_{13} \\ 
\ln \sigma_{14} \\ \ln \sigma - \ln \sigma_{14} \\ \ln \sigma - \ln \sigma_{13} \\ 
\ln \sigma - \ln \sigma_{12} 
\end{array} \right)
\end{equation}
Then the application of the Rouch\'{e}-Fr\"{o}benius theorem shows that system 
(\ref{syslin}) has a unique solution for $\{ \sigma_{1}, \sigma_{2}, \sigma_{3}, \sigma_{4} \}$ and the result is demonstrated. \hfill Q.E.D.

\mbox{}

Therefore notice that in Case I, Lemma 3.5 can be used to assume that all the $\sigma_{ij}>0$. Moreover, Lemma 3.6 can also be employed to write $\sigma_{ij}= \sigma_i \sigma_j$ in every case. Then from (\ref{red1}) we have the following type of Poisson matrix:
\begin{equation}
\label{red11}
	J'_{ij}(y)= \sigma _{i}\sigma _{j} \eta '(y)  \sum_{k,l=1}^4 \epsilon _{ijkl} 
	\phi '_l (y_l) \:\: , \:\:\: i,j = 1, \ldots , 4 
\end{equation}
with $\sigma_i>0$ for $i=1, \ldots ,4$. We can now state:

\mbox{}

{\em Lemma 3.7: \/} For an open set $\Omega \subset I \!\! R^4$, assume that ${\cal J} \in 
\Theta ( \Omega )$ is equivalent after transformation (\ref{yxdiff}) to a Poisson structure 
${\cal J'}$ of the form (\ref{red11}) defined in $y( \Omega ) = \Omega ' \subset I \!\! R^4$ and such that $\sigma_{i} > 0$ for $i= 1, \ldots ,4$. Then a complete set of independent Casimir invariants of such Poisson structure ${\cal J'}$ which are globally defined in $\Omega '$ and $C^2( \Omega ')$ is given by: 
\begin{equation}
\label{c11}
	C_1 (y) = \sigma_2 \sigma_3 \sigma_4 y_1 + \sigma_1 \sigma_3 \sigma_4 y_2 +
	\sigma_1 \sigma_2 \sigma_4 y_3 + \sigma_1 \sigma_2 \sigma_3 y_4
\end{equation}
\begin{equation}
\label{c12}
	C_2 (y) = \sigma_1 \sigma_2 \sigma_3 \sigma_4 \sum_{i=1}^4 
	\int \frac{\phi_i (y_i)}{\sigma_i} \mbox{d}y_i
\end{equation}

\mbox{}

{\em Proof of Lemma 3.7: \/} It is an application of the Pfaffian method.$^{26\:}$ 
\hfill Q.E.D.

\mbox{}

We can then proceed to the reduction to the Darboux canonical form in Case I. For this, consider the following change of variables globally diffeomorphic in $\Omega '$:
\begin{equation}
\label{tr12}
	\left\{ z_1=y_1 \: , \:\: z_2=y_2 \: , \:\: z_3=C_1(y) \: , \:\: z_4=C_2(y) \right\}
\end{equation}
where $C_1(y)$ and $C_2(y)$ are those in (\ref{c11}) and (\ref{c12}). When the transformation rule (\ref{jdiff}) is applied for (\ref{tr12}) to matrix (\ref{red11}) the result is: 
\begin{equation}
\label{darboux1}
	{\cal J}''(z) = \eta ''(z) \cdot 
	\left( \begin{array}{ccccc} 0 & 1 & 0 & 0 \\ -1 & 0 & 0 & 0 \\
	0 & 0 & 0 & 0 \\ 0 & 0 & 0 & 0 \end{array} \right)
\end{equation}
which is defined in $\Omega '' = z(\Omega ')$, and where $\eta '' (z)= \sigma_1 \sigma_2 \eta '(y(z)) ( \phi '_4(y(z))-\phi '_3(y(z)))$. To conclude, the reduction to the Darboux canonical form is achieved making use of Definition 3.1 to perform a time reparametrization of the form (\ref{ntt}), namely $\mbox{d} \tau = \eta ''(z) \mbox{d} t$, where $\tau$ is the new time and $\eta ''(z)$ is clearly nonvanishing in $\Omega ''$ and $C^1(\Omega '')$. According to (\ref{4dpos}) and (\ref{4dposntt}) the result is a new Poisson system with Darboux-type structure matrix:
\begin{equation}
	\label{jdarb4}
	{\cal J}_D =  \left( \begin{array}{cccc} 
	0 & 1 & 0 & 0 \\ -1 & 0 & 0 & 0 \\ 0 & 0 & 0 & 0 \\ 0 & 0 & 0 & 0 \end{array} \right)
\end{equation}
The reduction is thus globally completed in Case I.

\item[{\rm CASE II:}] $\sigma_{ij} = 0$ for some pair $(i,j)$, $i \neq j$. Again matrix 
(\ref{red1}) is our starting point. Now notice that $\sigma =0$ and as a consequence of conditions (\ref{condsig}) we actually have $\sigma _{ij}=0$ for at least three of the six pairs $(i,j)$, with $i<j$, $1 \leq i \leq 3$, $2 \leq j \leq 4$. This leads to eight possible subcases:
\begin{equation}
\label{iiabcases}
	\begin{array}{rl}
   \{ \: (II.A.1: \sigma _{14}= \sigma _{24}= \sigma _{34}=0) , &
	   (II.A.2: \sigma _{12}= \sigma _{13}= \sigma _{14}=0) , \\ \\
	   (II.A.3: \sigma _{12}= \sigma _{23}= \sigma _{24}=0) , &
	   (II.A.4: \sigma _{13}= \sigma _{23}= \sigma _{34}=0) , \\ \\
   	   (II.B.1: \sigma _{13}= \sigma _{14}= \sigma _{34}=0) , &
	   (II.B.2: \sigma _{12}= \sigma _{13}= \sigma _{23}=0) , \\ \\
	   (II.B.3: \sigma _{12}= \sigma _{14}= \sigma _{24}=0) , &
	   (II.B.4: \sigma _{23}= \sigma _{24}= \sigma _{34}=0)  \: \}
	\end{array}
\end{equation}
As it can be seen, these subcases are grouped in two different four-member sets (II.A and II.B). The four members of each set present analogous symplectic structures and similar reduction procedures to Darboux form. Let us start with the II.A possibilities:

\mbox{}

{\em Lemma 3.8: \/} For an open set $\Omega \subset I \!\! R^4$, assume that ${\cal J} \in 
\Theta ( \Omega )$ is equivalent after transformation (\ref{yxdiff}) to a Poisson structure 
${\cal J'}$ of the form (\ref{red1}) defined in $y( \Omega ) = \Omega ' \subset I \!\! R^4$ and corresponding to one of the subcases II.A.1 to II.A.4 in (\ref{iiabcases}). Then a complete set of independent Casimir invariants of such Poisson structure ${\cal J'}$ which are globally defined in $\Omega '$ and $C^2( \Omega ')$ is, respectively:
\begin{eqnarray*}
	II.A.1: & C_1 (y) = & y_4 \\
	        & C_2 (y) = & \sigma_{23} \int \phi_1 (y_1) \mbox{d}y_1 + 
		\sigma_{13} \int \phi_2 (y_2) \mbox{d}y_2 + 
		\sigma_{12} \int \phi_3 (y_3) \mbox{d}y_3 - \\ 
		  & & ( \sigma_{23} y_1 + \sigma_{13} y_2 + \sigma_{12} y_3 ) \phi_4(y_4) \\
	II.A.2: & C_1 (y) = & y_1 \\
	        & C_2 (y) = & \sigma_{34} \int \phi_2 (y_2) \mbox{d}y_2 + 
		\sigma_{24} \int \phi_3 (y_3) \mbox{d}y_3 + 
		\sigma_{23} \int \phi_4 (y_4) \mbox{d}y_4 - \\ 
		  & & ( \sigma_{34} y_2 + \sigma_{24} y_3 + \sigma_{23} y_4 ) \phi_1(y_1) \\
	II.A.3: & C_1 (y) = & y_2 \\
	        & C_2 (y) = & \sigma_{34} \int \phi_1 (y_1) \mbox{d}y_1 + 
		\sigma_{14} \int \phi_3 (y_3) \mbox{d}y_3 + 
		\sigma_{13} \int \phi_4 (y_4) \mbox{d}y_4 - \\ 
		  & & ( \sigma_{34} y_1 + \sigma_{14} y_3 + \sigma_{13} y_4 ) \phi_2(y_2) \\
	II.A.4: & C_1 (y) = & y_3 \\
	        & C_2 (y) = & \sigma_{24} \int \phi_1 (y_1) \mbox{d}y_1 + 
		\sigma_{14} \int \phi_2 (y_2) \mbox{d}y_2 + 
		\sigma_{12} \int \phi_4 (y_4) \mbox{d}y_4 - \\ 
		  & & ( \sigma_{24} y_1 + \sigma_{14} y_2 + \sigma_{12} y_4 ) \phi_3(y_3) 
\end{eqnarray*}

\mbox{}

{\em Proof of Lemma 3.8: \/} It is similar to the one of Lemma 3.7. \hfill Q.E.D.

\mbox{}

We carry out now the reduction to the Darboux canonical form for subcase II.A. For the sake of conciseness this will be done for the first possibility II.A.1, since the procedure is entirely analogous for the remaining situations II.A.2 to II.A.4. Thus for II.A.1 the following change of variables globally diffeomorphic in $\Omega '$ is defined:
\begin{equation}
\label{tr2a1}
	\left\{ v_1=y_1 \: , \:\: v_2=y_2 \: , \:\: v_3=C_2(y) \: , \:\: v_4=C_1(y) \right\}
\end{equation}
where $C_1(y)$ and $C_2(y)$ are those in Lemma 3.8 for subcase II.A.1 and according to hypothesis (c) of Theorem 2.1 it is assumed $\sigma _{12} \neq 0$ without loss of generality. Applying (\ref{jdiff}) and (\ref{tr2a1}) to such structure matrix it is again obtained a Poisson structure of the form ${\cal J''}(v) = \eta ''(v) \cdot {\cal J}_D$ defined in $v(\Omega ')$, where now $\eta '' (v)= \sigma_{12} \eta '(y(v)) ( \phi '_4(y(v))-\phi '_3(y(v)))$ and ${\cal J}_D$ is given in (\ref{jdarb4}). The reduction is concluded by means of a time reparametrization (\ref{ntt}) of the form $\mbox{d} \tau = \eta ''(v) \mbox{d} t$, where $\eta ''(v)$ is nonvanishing in $v( \Omega ')$ and $C^1(v( \Omega '))$. The result is thus a new Poisson system with structure matrix (\ref{jdarb4}) and the reduction is globally completed.

Consider next subcases II.B in (\ref{iiabcases}). For each of them both generic and nongeneric possibilities must be distinguished, according to the following definition: 

\mbox{}

{\em Definition 3.9: \/} Given a Poisson structure of the kind (\ref{red1}) characterized in Lemma 3.3 and corresponding to one of the subcases II.B.1 to II.B.4 in (\ref{iiabcases}), such structure will be called generic if only three of the six constants $\sigma _{ij}$ vanish, for $i<j$, $1 \leq i \leq 3$, $2 \leq j \leq 4$, while if four or five of such constants are zero the same type of structures will be termed nongeneric.

\mbox{}

Obviously the case in which all constants $\sigma_{ij}$ vanish is excluded due to condition (c) of Theorem 2.1. Now the generic II.B subcases will be treated first. For them we have the following result:

\mbox{}

{\em Lemma 3.10: \/} For an open set $\Omega \subset I \!\! R^4$, assume that ${\cal J} \in 
\Theta ( \Omega )$ is equivalent after transformation (\ref{yxdiff}) to a Poisson structure 
${\cal J'}$ of the form (\ref{red1}) defined in $y( \Omega ) = \Omega ' \subset I \!\! R^4$ and corresponding to one of the generic subcases II.B.1 to II.B.4 in (\ref{iiabcases}). Then a complete set of independent Casimir invariants of such Poisson structure ${\cal J'}$ which are globally defined in $\Omega '$ and $C^2( \Omega ')$ is, respectively:
\begin{eqnarray*}
	II.B.1: & C_1 (y) = & \sigma_{23} \sigma_{24} y_1 + \sigma_{12} \sigma_{24} y_3 +
			\sigma_{12} \sigma_{23} y_4 \\
	        & C_2 (y) = & \sigma_{23} \sigma_{24}  \int \phi_1 (y_1) \mbox{d}y_1 + 
		\sigma_{12} \sigma_{24} \int \phi_3 (y_3) \mbox{d}y_3 + 
		\sigma_{12} \sigma_{23} \int \phi_4 (y_4) \mbox{d}y_4  \\ 
	II.B.2: & C_1 (y) = & \sigma_{24} \sigma_{34} y_1 + \sigma_{14} \sigma_{34} y_2 +
			\sigma_{14} \sigma_{24} y_3 \\
	        & C_2 (y) = & \sigma_{24} \sigma_{34}  \int \phi_1 (y_1) \mbox{d}y_1 + 
		\sigma_{14} \sigma_{34} \int \phi_2 (y_2) \mbox{d}y_2 + 
		\sigma_{14} \sigma_{24} \int \phi_3 (y_3) \mbox{d}y_3  \\ 
	II.B.3: & C_1 (y) = & \sigma_{23} \sigma_{34} y_1 + \sigma_{13} \sigma_{34} y_2 +
			\sigma_{13} \sigma_{23} y_4 \\
	        & C_2 (y) = & \sigma_{23} \sigma_{34}  \int \phi_1 (y_1) \mbox{d}y_1 + 
		\sigma_{13} \sigma_{34} \int \phi_2 (y_2) \mbox{d}y_2 + 
		\sigma_{13} \sigma_{23} \int \phi_4 (y_4) \mbox{d}y_4  \\ 
	II.B.4: & C_1 (y) = & \sigma_{13} \sigma_{14} y_2 + \sigma_{12} \sigma_{14} y_3 +
			\sigma_{12} \sigma_{13} y_4 \\
	        & C_2 (y) = & \sigma_{13} \sigma_{14}  \int \phi_2 (y_2) \mbox{d}y_2 + 
		\sigma_{12} \sigma_{14} \int \phi_3 (y_3) \mbox{d}y_3 + 
		\sigma_{12} \sigma_{13} \int \phi_4 (y_4) \mbox{d}y_4 
\end{eqnarray*}

\mbox{}

{\em Proof of Lemma 3.10: \/} It is similar to the one of Lemma 3.7. \hfill Q.E.D.

\mbox{}

Regarding the reduction to the Darboux canonical form for the generic II.B subcases, possibility II.B.1 will be the only one explicitly considered, since again the procedure is completely analogous for the other cases II.B.2 to II.B.4. Then for II.B.1 (generic) the transformation globally diffeomorphic in $\Omega '$ to be performed is:
\begin{equation}
\label{tr2b1}
	\left\{ w_1=y_1 \: , \:\: w_2=y_2 \: , \:\: w_3=C_1(y) \: , \:\: w_4=C_2(y) \right\}
\end{equation}
where $C_1(y)$ and $C_2(y)$ are those in Lemma 3.10 for II.B.1. Once (\ref{tr2b1}) is defined, the rest of the reduction for the generic II.B.1 case is entirely similar to that of subcase II.A.1.

The only remaining situations are the nongeneric II.B subcases. The results to be presented are completely analogous for the four possibilities II.B.1 to II.B.4, and consequently we shall only deal explicitly with II.B.1 for the sake of brevity. For this, notice that there are two possible nongeneric situations for II.B.1:
\begin{description}
\item[{\rm II.B.1.a:}] One of $\{ \sigma_{12}, \sigma_{23}, \sigma_{24} \}$ vanishes. These three subcases are retrieved as particular instances of the II.A cases already analyzed, in such a way that the complete set of independent Casimir invariants and the reduction to the Darboux canonical form are also obtained as particular results of the ones given for II.A. Specifically, we may have: 
	\begin{description}
	\item $\bullet \:\:\: \sigma_{12}=0$: Such matrix is a particular case of II.A.2 in which 
		$\sigma_{34}=0$.
	\item $\bullet \:\:\: \sigma_{23}=0$: This is a particular case of II.A.4 with 
		$\sigma_{14}=0$.
	\item $\bullet \:\:\: \sigma_{24}=0$: It is a particular case of II.A.1 with 
		$\sigma_{13}=0$.
	\end{description}
\item[{\rm II.B.1.b:}] Two of $\{ \sigma_{12}, \sigma_{23}, \sigma_{24} \}$ vanish. Then the Casimir invariants are apparent and only a time reparametrization remains in order to reduce the Poisson system to Darboux form.
\end{description}
The classification is similar for the nongeneric II.B.2 to II.B.4 possibilities. Case II is thus concluded.
\end{description}

The demonstration of Theorem 3.2 is therefore complete. \hfill Q.E.D.

\mbox{}

Thus not only the Poisson structures considered but also their possible kinds of Casimir invariants and global reductions to the Darboux canonical form are completely characterized after the previous results. Once the main properties have been considered in detail, it is interesting to put in perspective the family just analyzed, as far as it is closely related to other Poisson structures reported in the literature. This is the aim of the next part of the work.

\mbox{}

\begin{flushleft}
{\bf IV. EXAMPLES AND RELATIONSHIP WITH OTHER SOLUTIONS}
\end{flushleft}

In this section the relationship of the family of solutions investigated with some other well-known Poisson structures is briefly explored. This is useful not only because the family of form (\ref{sol}) characterized in Theorem 2.1 provides a generalization of other structures or families of structures to be mentioned, but also because pointing up the intersections among different families should be helpful for future investigations regarding the Jacobi equations. Additionally, such illustrations provide interesting examples of the solutions analyzed throughout the paper.

Consider first the particular case of members of $\Theta ( \Omega )$ for which functions $\eta (x)$ and $\phi_i(x_i)$ $(i=1, \ldots ,4)$ have constant values. The result is always a separable Poisson structure,$^{31\:}$ namely a structure matrix of the form $J_{ij}=a_{ij} \psi_i (x_i) \psi_j (x_j) $, where the $a_{ij}$ are real constants that constitute the entries of a skew-symmetric matrix $A=(a_{ij})$, and the $\psi_i(x_i)$ are nonvanishing $C^1 ( \Omega )$ functions. Recall that separable matrices are always solutions of the Jacobi equations 
(\ref{sksym}-\ref{jac}) independently of the dimension of the Poisson manifold.$^{31\:}$ There are several interesting kinds of Poisson systems for which separable structures are natural in general dimension $n$, and consequently in the specific case of dimension $n=4$. This is the case of Poisson models arising in the domain of population dynamics (for either Lotka-Volterra$^{11\:}$ or generalized Lotka-Volterra$^{8\:}$ systems), plasma models$^{19\:}$ and systems such as the Toda and relativistic Toda lattices.$^{15\:}$ The interested reader is referred to the primary reference for further examples and the full details regarding issues such as the determination of the Casimir invariants and the reduction to the Darboux canonical form for separable Poisson structures.$^{31\:}$ Note in addition that according to Proposition 2.3 the structures belonging to $\Theta ( \Omega )$ have constant rank of value 2 everywhere in $\Omega$, while the rank of a separable matrix is the rank of $A$. Then it is interesting to remark that the particular case in which $\eta$ and $\phi_i$ ($i=1, \ldots ,4$) are constant does not comprise all possible four-dimensional separable matrices but only separable structures of rank two, thus illustrating an intersection between two known families of Poisson structures. 

As a second example, consider the limit case in which the functions $\psi_4(x_4) = 
\phi_4(x_4)=0$ are considered in (\ref{matrform}). In the resulting Poisson structure, it is clear that $x_4$ is a Casimir function. Then if a reduction is carried out to the symplectic leaf $x_4 = c$, the outcome is the 3-d Poisson structure of matrix:
\begin{equation}
\label{3dex}
	{\cal J}_{[3d]} = \tilde{\eta} \cdot
	\left( \begin{array}{ccc}
	 0 & \psi_1 \psi_2 \tilde{\phi}_3 & - \psi_1 \psi_3 \tilde{\phi}_2 \\
	- \psi_1 \psi_2 \tilde{\phi}_3 & 0 & \psi_2 \psi_3 \tilde{\phi}_1 \\
	\psi_1 \psi_3 \tilde{\phi}_2 & - \psi_2 \psi_3 \tilde{\phi}_1 & 0 \\
	\end{array} \right)
\end{equation}
where $\tilde{\eta} (x_1,x_2,x_3) = \eta (x_1,x_2,x_3,c)$ and $\tilde{\phi}_i = \sigma_{jk} \phi_i$ for $i=1,2,3$, where $(i,j,k)$ denotes an arbitrary permutation of $(1,2,3)$. Dropping the tildes for the sake of clarity, the resulting structures can also be expressed as: 
\begin{equation}
\label{3dex2}
	({\cal J}_{[3d]})_{ij}(x_1,x_2,x_3)= \eta (x_1,x_2,x_3) \psi_i(x_i) \psi_j(x_j) 
	\sum_{k=1}^3 \epsilon_{ijk} \phi_k (x_k) \:\: , \:\:\:\:\: i,j=1,2,3 
\end{equation}
Poisson structures of the form (\ref{3dex}-\ref{3dex2}) have been studied in detail in the literature,$^{38\:}$ and actually they comprise as particular cases very different Poisson matrices employed before in several domains, including the Euler top,$^{2\:}$ the Kermack-McKendrick model,$^{10,37\:}$ certain integrable cases of the Lorenz system,$^{20\:}$ population models such as those of Lotka-Volterra$^{9,11,37\:}$ and generalized 
Lotka-Volterra$^{8\:}$ types, the Maxwell-Bloch equations,$^{27\:}$ the Rabinovich system,$^{20\:}$ or the RTW interaction equations.$^{20\:}$ A discussion of these particular instances as well as an analysis of structures (\ref{3dex}-\ref{3dex2}) including their symplectic structure, Casimir invariants and construction of the Darboux coordinates are present in the aforementioned reference.$^{38\:}$ Such family is also interesting from the point of view of the separable structures considered in the first part of this section, since it is evident that all three-dimensional separable structures are particular cases of (\ref{3dex2}).

It can be thus appreciated how the identification of the solutions characterized in Theorem 2.1 leads to the establishment of some new links among different families of Poisson structures.

\mbox{}

\begin{flushleft}
{\bf V. FINAL REMARKS}
\end{flushleft}

Every new contribution to the study of skew-symmetric solutions of the Jacobi equations tends to provide a more general perspective of the field of finite-dimensional Poisson structures. Typical features of this fact can be appreciated in the previous analysis. Not only the identification of new finite-dimensional Poisson structures constitutes in itself a relevant problem from the point of view of mathematical physics, but in addition this knowledge provides a richer framework for the fundamental problem of recasting a given differential flow into a Poisson system, whenever possible. Additionally, it is worth noting that the characterization of a sufficiently general solution family often allows the conceptual and operational unification of diverse Poisson structures and systems previously well-known but unrelated, which can hereafter be regarded from a more general and economic standpoint. Examples of this have been given in Section IV. In particular, in such sense it is physically interesting to identify the Casimir invariants and to develop the reduction procedure to the Darboux canonical form for the new solution families. These are features of special relevance when they can be globally achieved, thus providing an additional instance of a result that goes beyond the {\em a priori \/} scope of Darboux theorem and has been reported only in a limited number of cases. This kind of results suggests that the direct investigation of the Jacobi equations constitutes a fruitful line of research not only for classification purposes but also for the detailed analysis of Poisson structures, not to mention its mathematical interest as an example of nonlinear system of PDEs. Additionally to these considerations, it is worth recalling that dimension three is the simplest nontrivial case for the analysis of the Jacobi equations and has consequently been studied in much more detail than higher dimensions, as discussed in the Introduction. On the other hand, Jacobi equations (\ref{jac}) become increasingly complicated as dimension grows. This explains the relative scarcity of results for dimensions four and higher. Certainly, a complete knowledge of the skew-symmetric solutions of the Jacobi equations is still far, but nevertheless the investigation of the problem seems to be a unavoidable issue for a better understanding of finite-dimensional Poisson structures, and therefore of the scope of Hamiltonian dynamics. 

\pagebreak
\begin{flushleft}
{\bf References and notes}
\end{flushleft}
\begin{description}
\addtolength{\itemsep}{-0.2cm}
\item[$^1$] A. Lichnerowicz, J. Diff. Geom. {\bf 12}, 253 (1977); A. Weinstein, J. Diff. Geom. {\bf 18}, 523 (1983).  
\item[$^2$] P. J. Olver, {\em Applications of Lie Groups to Differential Equations}, 2nd ed. 
	(Springer-Verlag, New York, 1993).  
\item[$^{3}$] P. J. Morrison, Rev. Mod. Phys. {\bf 70}, 467 (1998).  
\item[$^{4}$] R. D. Hazeltine, D. D. Holm and P. J. Morrison, J. Plasma Phys. {\bf 34}, 103 
	(1985); D. D. Holm, Phys. Lett. A {\bf 114}, 137 (1986); P. J. Morrison and J. M. Greene, 
	Phys. Rev. Lett. {\bf 45}, 790 (1980). 
\item[$^{5}$] J. E. Marsden, R. Montgomery, P. J. Morrison and W. B. Thompson, Ann. Phys. 	(N.Y.) {\bf 169}, 29 (1986).
\item[$^{6}$] I. E. Dzyaloshinskii and G. E. Volovick, Ann. Phys. (N.Y.) {\bf 125}, 67 
	(1980).  
\item[$^7$] L. Cair\'{o} and M. R. Feix, J. Phys. A {\bf 25}, L1287 (1992). 
\item[$^{8}$] B. Hern\'{a}ndez--Bermejo and V. Fair\'{e}n, J. Math. Phys. {\bf 39}, 6162 
	(1998); B. Hern\'{a}ndez--Bermejo and V. Fair\'{e}n, J. Math. Anal. Appl. {\bf 256}, 242 
	(2001).
\item[$^{9}$] Y. Nutku, Phys. Lett. A {\bf 145}, 27 (1990).  
\item[$^{10}$] Y. Nutku, J. Phys. A {\bf 23}, L1145 (1990).  
\item[$^{11}$] M. Plank, J. Math. Phys. {\bf 36}, 3520 (1995); M. Plank, SIAM (Soc. Ind. Appl. 	Math.) J. Appl. Math. {\bf 59}, 1540 (1999).
\item[$^{12}$] M. Plank, Nonlinearity {\bf 9}, 887 (1996). 
\item[$^{13}$] F. Haas, J. Phys. A: Math. Gen. {\bf 35}, 2925 (2002).
\item[$^{14}$] K. Marciniak and S. Rauch-Wojciechowski, J. Math. Phys. {\bf 39}, 5292 (1998). 
\item[$^{15}$] P. A. Damianou, J. Math. Phys. {\bf 35}, 5511 (1994). 
\item[$^{16}$] S. A. Hojman, J. Phys. A {\bf 24}, L249 (1991); S. A. Hojman, J. Phys. A {\bf 	29}, 667 (1996); C. A. Lucey and E. T. Newman, J. Math. Phys. {\bf 29}, 2430 (1988); V. 	Perlick, J. Math. Phys. {\bf 33}, 599 (1992). 
\item[$^{17}$] R. G. Littlejohn, J. Math. Phys. {\bf 20}, 2445 (1979); R. G. Littlejohn, J. 	Math. Phys. {\bf 23}, 742 (1982); J. R. Cary and R. G. Littlejohn, Ann. Phys. (N.Y.) {\bf 	151}, 1 (1983).
\item[$^{18}$] D. David, D. D. Holm and M. V. Tratnik, Phys. Rep. {\bf 187}, 281 (1990).
\item[$^{19}$] G. Picard and T. W. Johnston, Phys. Rev. Lett. {\bf 48}, 1610 (1982). 
\item[$^{20}$] J. Goedert, F. Haas, D. Hua, M. R. Feix and L. Cair\'{o}, J. Phys. A {\bf 27}, 	6495 (1994). 
\item[$^{21}$] F. Haas and J. Goedert, Phys. Lett. A {\bf 199}, 173 (1995); 
	B. Hern\'{a}ndez--Bermejo and V. Fair\'{e}n, Phys. Lett. A {\bf 234}, 35 (1997).  
\item[$^{22}$] G. B. Byrnes, F. A. Haggar and G. R. W. Quispel, Physica A {\bf 272}, 99 
	(1999).
\item[$^{23}$] R. I. McLachlan, Phys. Rev. Lett. {\bf 71}, 3043 (1993);
	R. I. McLachlan, G. R. W. Quispel and N. Robidoux, Phys. Rev. Lett. {\bf 81}, 2399 	(1998).
\item[$^{24}$] D. D. Holm, J. E. Marsden, T. Ratiu and A. Weinstein, Phys. Rep. {\bf 123}, 1 	(1985). 
\item[$^{25}$] J. C. Simo, T. A. Posbergh and J. E. Marsden, Phys. Rep. {\bf 193}, 279 (1990). 
\item[$^{26}$] B. Hern\'{a}ndez--Bermejo and V. Fair\'{e}n, Phys. Lett. A {\bf 241}, 148 	(1998); T. W. Yudichak, B. Hern\'{a}ndez--Bermejo and P. J. Morrison, Phys. Lett. A {\bf 	260}, 475 (1999). 
\item[$^{27}$] D. David and D. D. Holm, J. Nonlinear Sci. {\bf 2}, 241 (1992). 
\item[$^{28}$] P. J. Olver, Phys. Lett. A {\bf 148}, 177 (1990); P. Gao, Phys. Lett. A {\bf 	273}, 85 (2000); C. Gonera and Y. Nutku, Phys. Lett. A {\bf 285}, 301 (2001).
\item[$^{29}$] R. G. Littlejohn, J. Plasma Phys. {\bf 29}, 111 (1983); P. Crehan, Prog. 	Theor. Phys. Suppl. {\bf 110}, 321 (1992).
\item[$^{30}$] K. Ngan, S. Meacham and P. J. Morrison, Phys. Fluids {\bf 8}, 896 (1996).
\item[$^{31}$] B. Hern\'{a}ndez--Bermejo and V. Fair\'{e}n, Phys. Lett. A {\bf 271}, 258 
	(2000).
\item[$^{32}$] B. Hern\'{a}ndez--Bermejo, Phys. Lett. A {\bf 287}, 371 (2001).
\item[$^{33}$] S. Lie, {\em Theorie der Transformationsgruppen\/} (B. G. Teubner, Leipzig, 	1888). 
\item[$^{34}$] K. H. Bhaskara, Proc. Indian Acad. Sci. Math. Sci. {\bf 100}, 189 (1990). 
\item[$^{35}$] K. H. Bhaskara and K. Rama, J. Math. Phys. {\bf 32}, 2319 (1991). 
\item[$^{36}$] Z.-J. Liu and P. Xu, Lett. Math. Phys. {\bf 26}, 33 (1992). 
\item[$^{37}$] H. G\"{u}mral and Y. Nutku, J. Math. Phys. {\bf 34}, 5691 (1993). 
\item[$^{38}$] B. Hern\'{a}ndez--Bermejo, J. Math. Phys. {\bf 42}, 4984 (2001). 
\end{description}
\end{document}